\documentclass[12pt]{iopart}

\usepackage{epsfig}

\begin{document}

\title[Metamagnetic transition in EuFe$_2$As$_2$]{Metamagnetic transition in EuFe$_2$As$_2$ single crystals}

\author{Shuai Jiang, Yongkang Luo, Zhi Ren, Zengwei Zhu, Cao Wang, Xiangfan Xu, Qian Tao, Guanghan Cao\dag, and Zhu'an Xu\ddag}

\address{Department of Physics, Zhejiang University, Hangzhou 310027, People's Republic of China}
\ead{\dag ghcao@zju.edu.cn; \ddag zhuan@zju.edu.cn}

\begin{abstract}
We report the measurements of anisotropic magnetization and
magnetoresistance on single crystals of EuFe$_2$As$_2$, a parent
compound of ferro-arsenide high-temperature superconductor. Apart
from the antiferromagnetic (AFM) spin-density-wave transition at 186
K associated with Fe moments, the compound undergoes another
magnetic phase transition at 19 K due to AFM ordering of Eu$^{2+}$
spins ($J=S=7/2$). The latter AFM state exhibits metamagnetic
transition under magnetic fields. Upon applying magnetic field with
$H\parallel c$ at 2 K, the magnetization increases linearly to 7.0
$\mu_{B}$/f.u. at $\mu_{0}H$=1.7 T, then keeps at this value of
saturated Eu$^{2+}$ moments under higher fields. In the case of
$H\parallel ab$, the magnetization increases step-like to 6.6
$\mu_{B}$/f.u. with small magnetic hysteresis. A metamagnetic phase
was identified with the saturated moments of 4.4 $\mu_{B}$/f.u. The
metamagnetic transition accompanies with negative in-plane
magnetoresistance, reflecting the influence of Eu$^{2+}$ moments
ordering on the electrical conduction of FeAs layers. The results
were explained in terms of spin-reorientation and spin-reversal
based on an $A$-type AFM structure for Eu$^{2+}$ spins. The magnetic
phase diagram has been established.
\end{abstract}



\maketitle

\section{Introduction}
The discovery of high temperature superconductivity in
LnFeAsO$_{1-x}$F$_{x}$ (Ln=lanthanides)\cite{Hosono,Chen-Sm,WNL-Ce}
has stimulated intense research in the field of condensed matter
physics. The superconducting transition temperature has achieved 55
K or more by either high-pressure synthesis\cite{Ren-Sm,Kito} or
Th-doping strategy\cite{Wang-Th}. The key structural unit of the
superconductors is accepted as the antifluorite-type
[Fe$_2$As$_2$]$^{2-}$ layers. This point of view is manifested by
the observation of superconductivity up to $\sim$ 38 K in
Ba$_{1-x}$K$_x$Fe$_2$As$_2$\cite{BaFe2As2 SC},
Sr$_{1-x}$K$_x$Fe$_2$As$_2$\cite{SrFe2As2 SC,ChenXH},
Ca$_{1-x}$Na$_x$Fe$_2$As$_2$\cite{CaFe2As2 SC}, and
Li$_{1-x}$FeAs\cite{JCQ}, all of which contain similar
[Fe$_2$As$_2$]$^{2-}$ layers. Another important point is that the Fe
sublattice of the parent compound is antiferromagnetic (AFM) in the
ground state\cite{DaiPC,Huang}, and superconductivity is induced by
suppressing the AFM order through appropriate carrier doping.

EuFe$_2$As$_2$\cite{EuFeAs} belongs to the so-called "122" family
$A$Fe$_2$As$_2$ ($A$=Ba, Sr, Ca and Eu), and it stands out due to
the magnetic moments of Eu$^{2+}$. We have recently performed a
systematic physical property measurements on EuFe$_2$As$_2$
polycrystalline sample.\cite{Eu122-Ren} Very similar magnetic
transition related to Fe$_2$As$_2$ layers was revealed between
EuFe$_2$As$_2$ and SrFe$_2$As$_2$. By assuming that Eu$^{2+}$
moments are compatible with superconductivity, we had anticipated that
superconductivity might be realized by proper doping in
EuFe$_2$As$_2$ systems. As a matter of fact, superconductivity was
indeed obtained in Eu$_{0.5}$K$_{0.5}$Fe$_2$As$_2$ and
Eu$_{0.7}$Na$_{0.3}$Fe$_2$As$_2$, according to very recent
reports\cite{Jeevan,EuSC-Qi}.

Although the free Eu$^{2+}$ moments do not directly affect
superconductivity, study on the ordering of Eu$^{2+}$ moments may
shed light on the mechanism of high-temperature superconductivity in
iron arsenides. Our preceding work\cite{Eu122-Ren} indicated that
the magnetic ordering of Eu$^{2+}$ moments in EuFe$_2$As$_2$ was
very intriguing. While the Eu$^{2+}$ spins ($S$=7/2, $L$=0) order
antiferromagnetically below 19 K at zero field, the Curie-Weiss fit
of high-temperature magnetic susceptibility suggests ferromagnetic
interactions between the Eu$^{2+}$ spins. When applying magnetic
field, a metamagnetic transition was found around 0.65 T. To further
understand the intrinsic properties of this magnetically ordered
materials, we performed the measurements of anisotropic
magnetization and magnetoresistance on single crystals of
EuFe$_2$As$_2$. As a result, anisotropic metamagnetic transitions
were uncovered. What is more, the electrical conduction of FeAs
layers was found to be related to the magnetic state of Eu layers.

\section{Experimental details}

Single crystals of EuFe$_2$As$_2$ were grown using FeAs as the
self-flux similar to previous report\cite{CXH-SX}. FeAs was
presynthesized by reacting Fe powders with As shots in vacuum at 773
K for 6 hours and then 1030 K for 12 hours. Fresh Eu grains and FeAs
powders were thoroughly mixed in a molar ratio of 1:4. The mixture
was loaded into an alumina tube which was put into a quartz ampoule.
The sealed quartz ampoule was heated to 1053 K at a rate of 150 K/h,
holding for 10 hours. Subsequently, the temperature was raised to
1398 K in 3 hours, holding for 5 hours. The crystals were grown by
slow cooling to 1223 K at a rate of 2 K/h. Finally the quartz
ampoule was furnace-cooled to room temperature. Many shiny
plate-like crystals with the typical size of $1.5\times 1.5\times
0.1$ mm$^{3}$ were obtained.

X-ray diffraction (XRD) was performed using a D/Max-rA
diffractometer with Cu-K$_{\alpha}$ radiation and a graphite
monochromator. Fig. 1 shows the XRD pattern of EuFe$_2$As$_2$
crystals. Only ($00l$) reflections with even $l$ appear, indicating
that the $c$-axis is perpendicular to the crystal plate. The
$c$-axis was calculated as 12.11 {\AA}, consistent with our previous
measurement using polycrystalline samples\cite{Eu122-Ren}.

\begin{figure}
\center
\includegraphics{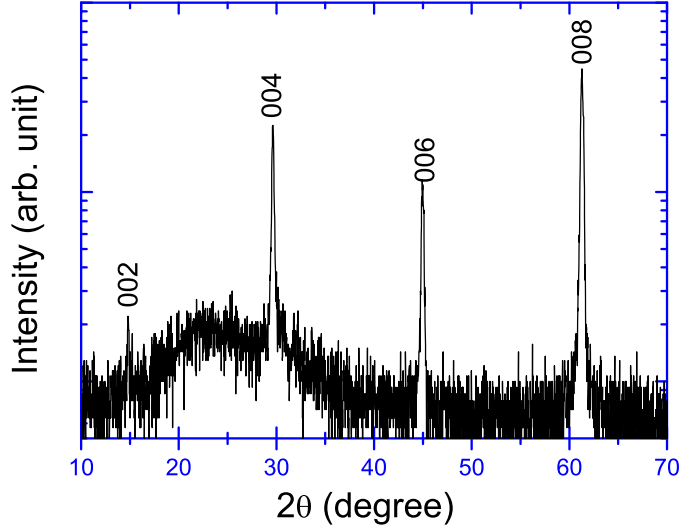}
\caption{X-ray multiple diffraction pattern for EuFe$_2$As$_2$
plate-like crystals lying on the sample holder. Note that the
logarithmic scale was employed for the intensity axis to verify the
sample quality. The hump around $2\theta$=25$^\circ$ is due to the
diffractions of glass sample holder.}
\end{figure}

Electrical resistivity was measured using a standard four-terminal
method under magnetic field up to 5 T. The dc magnetization was
measured on a Quantum Design magnetic property measurement system
(MPMS-5). The plate-like crystal was carefully mounted in a sample
holder, so that the applied field was basically perpendicular or
parallel to crystallographic $c$-axis. The deviation angle was
estimated to be less than 5$^{\circ}$.

\section{Results and discussion}
Figure 2 shows the temperature dependence of magnetic susceptibility
($\chi$) of EuFe$_2$As$_2$ crystals in two orientations of magnetic
field. At high temperatures ($T>50$ K), there is no difference
between $\chi_{ab}$ and $\chi_{c}$, indicating isotropic
susceptibility. In the range of 19 K $\leq T<$ 50 K, however, a
significant anisotropy in susceptibility (\emph{e. g.},
$\chi_{ab}/\chi_{c} =1.35$ at 19 K) shows up, suggesting an
anisotropic magnetic interaction. Below 19 K, $\chi_{ab}$ decreases
very sharply, while $\chi_{c}$ almost remains constant with
decreasing temperature, indicating a Neel transition. This
observation strongly suggests that the Eu$^{2+}$ moments align
within $ab$ planes, which is different from the previous proposal by
$^{151}$Eu M\"{o}ssbauer study\cite{Mossbuar}.

\begin{figure}
\center
\includegraphics[width=7cm]{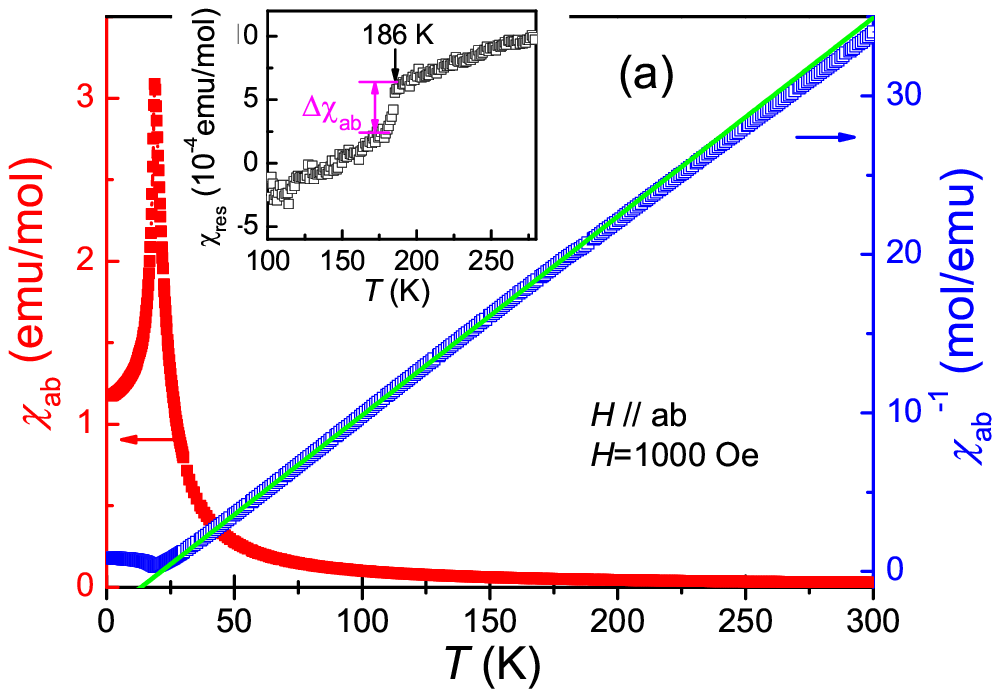}
\includegraphics[width=7cm]{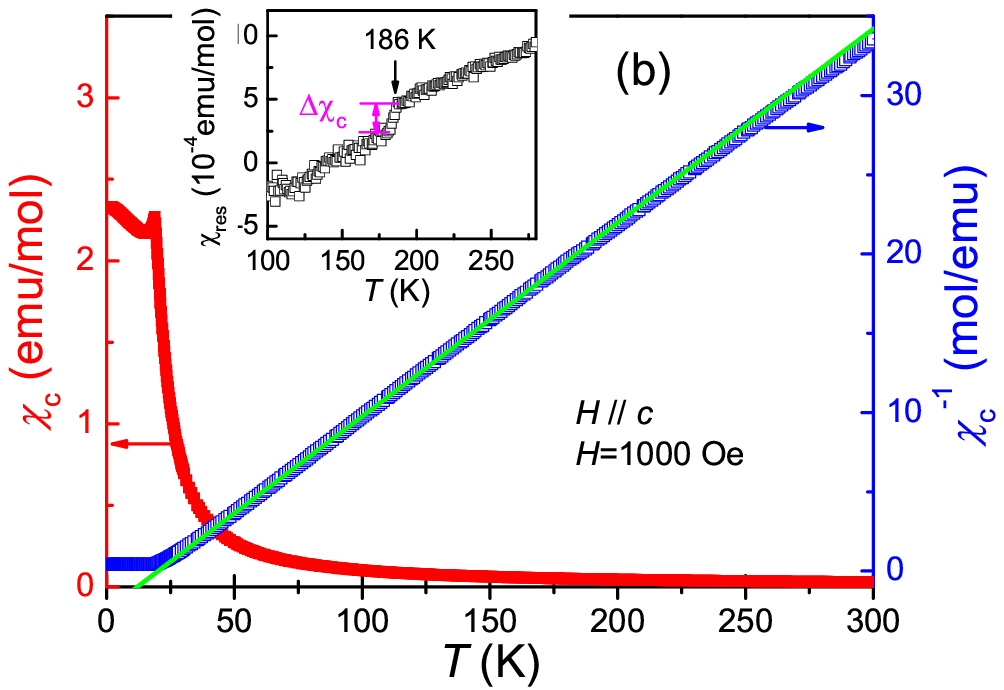}
\caption{Temperature dependence of magnetic susceptibility of
EuFe$_2$As$_2$ crystals with the magnetic field ($\mu_{0}H$=0.1T)
perpendicular (a) and parallel (b) to crystallographic $c$-axis. The
straight lines are guide to the eyes. Both insets show a drop in
$\chi$ at 186 K, after subtraction of the Curie-Weiss contribution
of Eu$^{2+}$ moments.}
\end{figure}

The high-temperature $\chi(T)$ data follows the extended Curie-Weiss
law,
\begin{equation}
\chi=\chi_{0}+\frac{C}{T+\theta},
\end{equation}
where $\chi_{0}$ is the temperature-independent term of the
susceptibility, $C$ the Curie constant and $\theta$ the Weiss
temperature. The fitted parameters and the derived effective
magnetic moments are listed in Table I. For both $H\parallel c$ and
$H\parallel ab$, the experimental value of Eu$^{2+}$ moments is
close to the theoretical value of $g\sqrt{S(S+1)}$=7.94 $\mu_{B}$
with $S=7/2$ and $g$=2. The Weiss temperature is negative,
indicating predominately ferromagnetic interaction among Eu$^{2+}$
spins. To reconcile the AFM ordering and the ferromagnetic
interaction, and considering the enhanced $\chi_{ab}$ just above the
Neel temperature, the Eu$^{2+}$ spins probably align
ferromagnetically within $ab$ planes, but antiferromagnetically
along the $c$-axis (see the inset of Fig. 5).
This magnetic structure of Eu sublattice resembles that of LaMnO$_3$,
which was called $A$-type antiferromagnetism\cite{LaMnO3}. A more
relevant example is RNi$_2$B$_2$C (R=Pr, Dy and Ho) whose magnetic
structure is also of $A$-type\cite{RNi2B2C}. Further experiments
such as neutron diffractions are needed to confirm this magnetic structure.

\begin{table}
\caption{\label{tabone}Magnetic parameters from the fitting of the
high-temperature (50 K $\sim$ 180 K) susceptibility data for
EuFe$_2$As$_2$ crystals using Eq.(1).}

\begin{indented}
\lineup
\item[]\begin{tabular}{@{}*{3}{l}}
\br Fitted Parameters&\0\0\0\0\0\0$H\parallel
ab$&\0\0\0\0\0\0$H\parallel c$\cr \mr$\chi_{0}$ (emu/mol) &
\0\0\0\0\0\0-0.00022 & \0\0\0\0\0\0-0.00082\cr $C$ (emu K/mol)
&\0\0\0\0\0\07.99 & \0\0\0\0\0\08.31\cr  $\theta$  (K)  &
\0\0\0\0\0\0-21.4 &\0\0\0\0\0\0 -19.7\cr $\mu_{eff}$
($\mu_{B}$/f.u.) & \0\0\0\0\0\07.97 & \0\0\0\0\0\08.13\cr \br
\end{tabular}
\end{indented}
\end{table}

After subtracting the above Curie-Weiss contribution, a small drop
in $\chi$ at 186 K can be found for both field orientations. This
anomaly in $\chi$ has been identified due to the AFM SDW
transition\cite{Eu122-Ren}, though the anomaly temperature is
somewhat lower than that of the polycrystalline sample.
$\Delta\chi_{ab}$ is significantly larger than $\Delta\chi_{c}$,
supporting that the Fe moments align within $ab$ planes in analogue
with that in other related iron arsenids revealed by the neutron
diffraction studies\cite{DaiPC,Huang}. In the SDW state, Fe$^{2+}$
moments order antiferromagnetically with a collinear
\emph{stripe-like} spin structure. Thus, the coupling between
Eu$^{2+}$ and Fe$^{2+}$ moments would be geometrically frustrated.
Besides, the energy scale of AFM coupling of Fe$^{2+}$ moments is
estimated to be much higher than the AFM interlayer coupling of
Eu$^{2+}$ moments. Therefore, the magnetic coupling between
Eu$^{2+}$ and Fe$^{2+}$ moments is negligible in the following
discussion.

\begin{figure}
\center
\includegraphics[width=7cm]{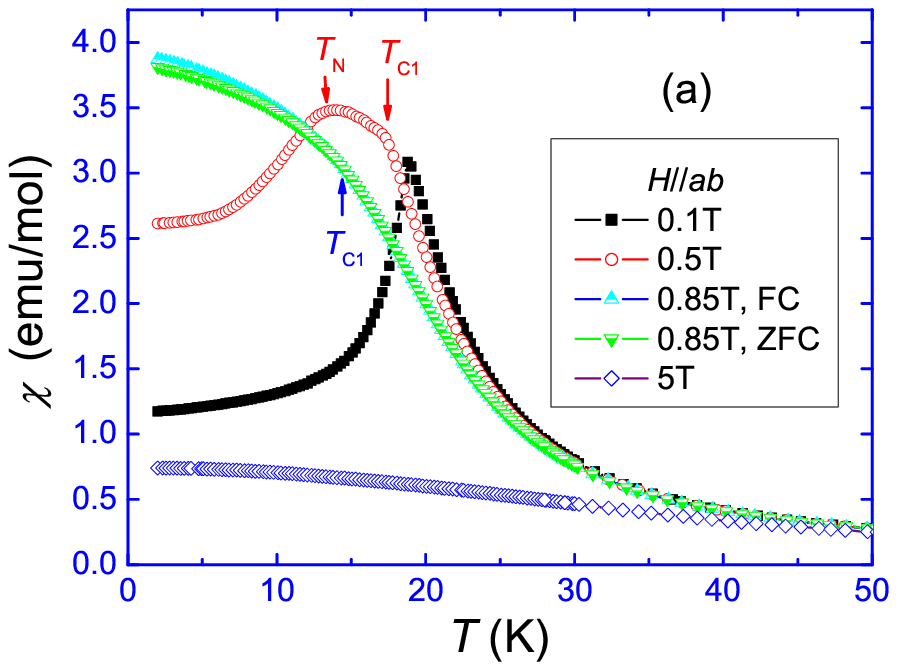}
\includegraphics[width=7cm]{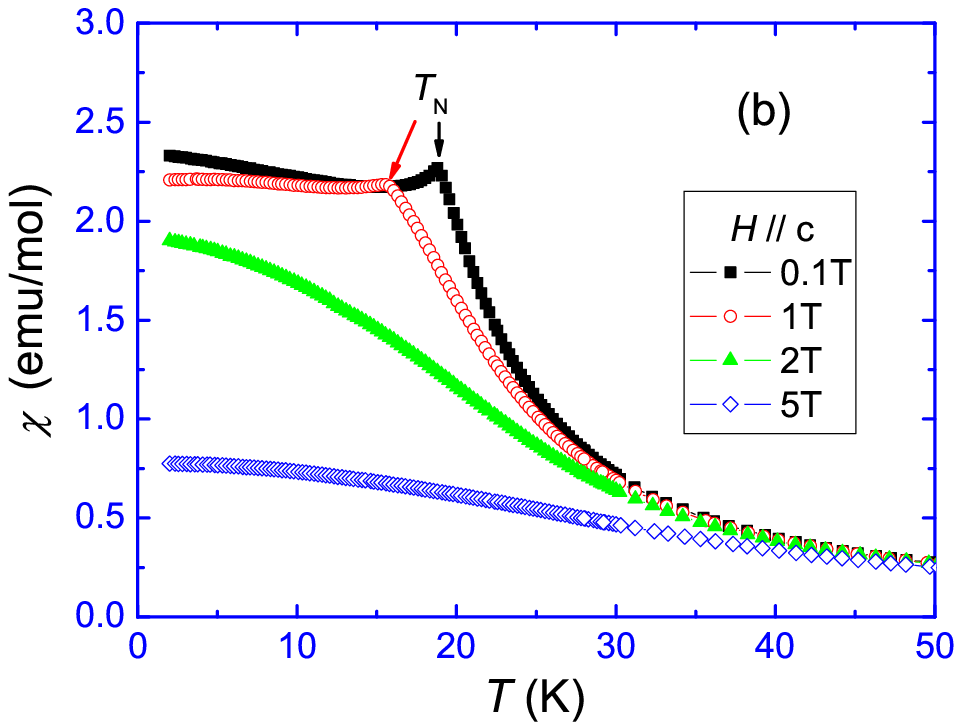}
\caption{Temperature dependence of magnetic susceptibility of
EuFe$_2$As$_2$ crystals under various magnetic fields. The magnetic
field is perpendicular (a) and parallel (b) to crystallographic
$c$-axis.}
\end{figure}

$A$-type antiferromagnetism often undergoes metamegnetic transition
under strong magnetic field because of relatively weak interlayer
AFM coupling. Fig. 3 shows the $\chi(T)$ curves under various
magnetic fields. At low magnetic field of $\mu_{0}H$=0.1T, AFM
transition takes place at 19 K. For $\mu_{0}H_{\parallel ab}=$ 0.5
T, however, successive magnetic transitions were observed. First, a
kink in $\chi$ appears at $T_{C1}$=17 K. Then, $\chi$ starts to drop
below $T_{N}$=13 K. At lower temperatures down to 2 K, there exists
impressively large residual susceptibility. When
$\mu_{0}H_{\parallel ab}$ is increased to 0.85 T, only one magnetic
transition can be distinguished. The transition has small magnetic
hysteresis, suggesting a kind of ferromagnetism. For $H\parallel c$,
the Neel temperature is decreased by the applied fields for
$\mu_{0}H_{\parallel c}< $2 T. When $\mu_{0}H_{\parallel c}\geq $2
T, the AFM transition was suppressed.

Figure 4 shows the field-dependent magnetization for EuFe$_2$As$_2$
crystals at various temperatures. At 50 K, which is well above the
Neel temperature $T_N$, the $M(H)$ curve is essentially linear. When
the temperature is close to $T_N$, a strong non-linearity in
magnetization can be seen. Below $T_N$, $M_{ab}$ first increases
almost linearly, then increases abruptly to a certain value
(depending on temperature), finally continues to increase to a
saturated value. Small magnetic hysteresis was identified. In the
case of $M_{c}$, no such step-like magnetization behavior with
magnetic hysteresis was observed. At 2 K, for example, $M_{c}$
increases linearly to 7.0 $\mu_{B}$/f.u. at $\mu_{0}H$=1.7 T, then
keeps at this value of saturated Eu$^{2+}$ moments ($M_{sat}=gS$=7.0
$\mu_{B}$/f.u. for $g$=2 and $S$=7/2) for higher fields. The linear
field dependence of $M_{c}$ is consistent with spin re-orientation,
since applied field rotates the moment gradually from $\perp c$ to
$\parallel c$.

\begin{figure}
\center
\includegraphics[width=6.8cm]{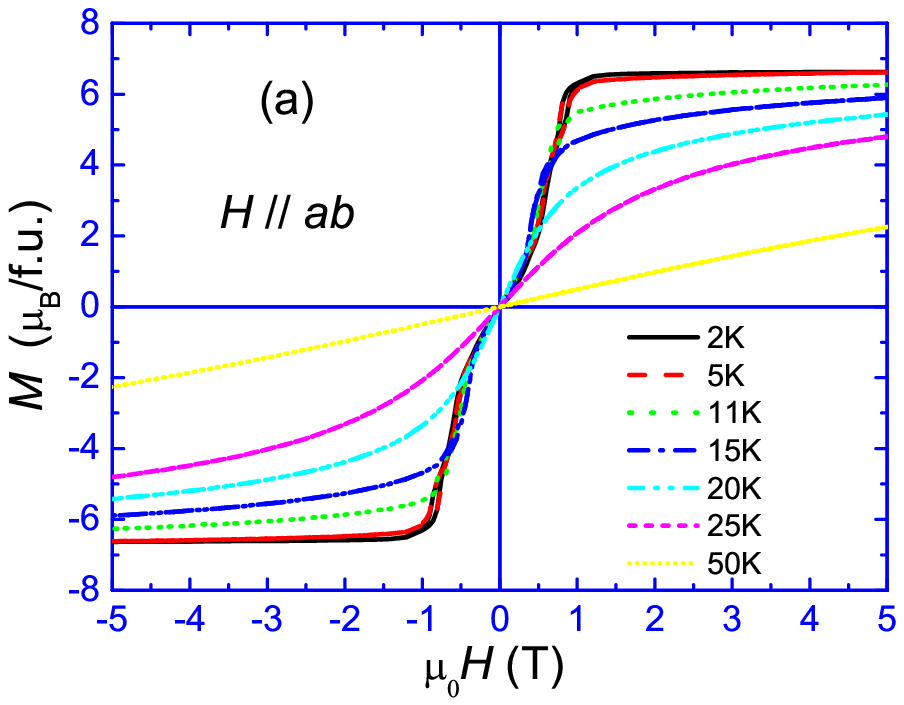}
\includegraphics[width=7cm]{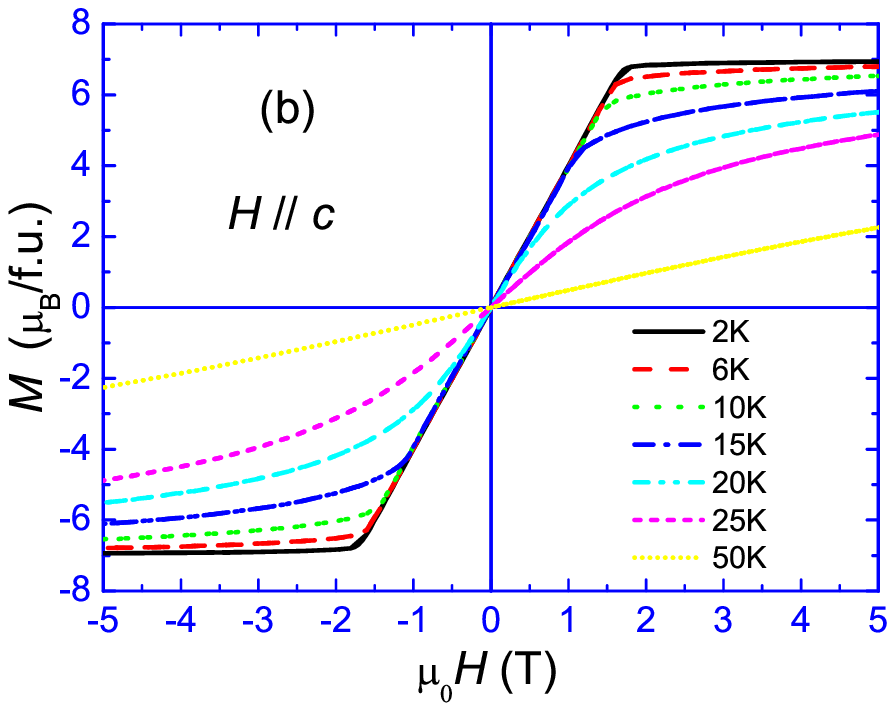}
\caption{Magnetic field dependence of magnetization of
EuFe$_2$As$_2$ crystals with the field perpendicular (a) and
parallel (b) to crystallographic $c$-axis.}
\end{figure}

To analyze the complex magnetization for $H\parallel ab$, an
expanded plot is shown in Fig. 5. The linear increase in $M_{ab}$
below 0.45 T probably corresponds to spin re-orientation. In the
field range of 0.5 T $< \mu_{0}H<$ 0.7 T, $M_{ab}$ increases rapidly
to 4.4 $\mu_{B}$/f.u. Because of the small magnetic hysteresis, the
rapid increase in $M$ above 0.45 T is unlikely due to a spin-flop
transition, and we ascribe it to a metamagnetic transition. For 0.7
T $< \mu_{0}H<$ 1.0 T, another ferromagnetic loop can be seen.
$M_{ab}$ finally saturates to 6.6 $\mu_{B}$/f.u. above 1.0 T. The
saturated moment is a little smaller than the expected value of 7.0
$\mu_{B}$/f.u., which is possibly due to the crystal field effect.
It is noted that the intermediate magnetization of 4.4
$\mu_{B}$/f.u. is just 2/3 of the saturated one. Therefore, we
propose a possible configuration for the intermediate metamagnetic
(MM) state: In every six sheets of Eu$^{2+}$, five of them have the
moment parallel to the external field, and the remained one has the
moment antiparallel to the applied field. We note that similar MM
phases were found in RNi$_2$B$_2$C system\cite{Detlefs}.

\begin{figure}
\center
\includegraphics[width=7cm]{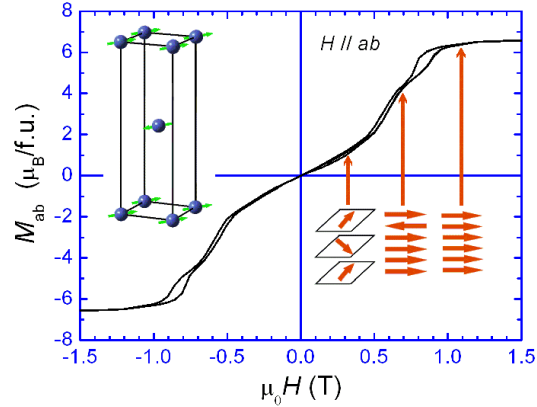}
\caption{Expanded $M-H$ plot for $H\parallel ab$ at 2 K. The insets
give the possible magnetic structure at zero field (upper left, each
ball represents Eu atom with spin 7/2), and the configuration of
magnetic polarization (lower right, each arrow represents the
magnetic moment in a Eu$^{2+}$ sheet).}
\end{figure}

Figure 6 shows the isothermal in-plane resistance ($R$) under the
applied field perpendicular or parallel to the crystallographic
$c$-axis. At 20 K, which is very close to $T_N$, the resistance
decreases gradually at low fields, and then almost remains unchanged
under higher fields. The negative magnetoresistance (MR) is ascribed
to the reduction of spin disorder scattering, since the paramagnetic
Eu$^{2+}$ spins tend to align along the external magnetic field. At
the temperature far below the $T_N$ (\emph{e.g.}, at 2 K) in which
Eu$^{2+}$ spins order antiferromagnetically, the resistance first
decreases to a minimum, then increases almost linearly. The turning
point at $H_{C3}$ corresponds to the onset of the magnetic
saturation in $M(H)$ curves. The negative MR below $H_{C3}$ suggests
that the AFM-ordered Eu$^{2+}$ spins scatter the charge transport in
FeAs layers, similar to the well known giant magnetoresistance
observed in magnetic multilayers\cite{GMR}. The increase of MR above
$H_{C3}$ (where Eu$^{2+}$ spins order ferromagnetically) reflects
the intrinsic property of the SDW state. In fact, positive MR was
observed at low temperature for LaOFeAs, which was explained in
terms of the suppression of SDW order by external magnetic
field\cite{WNL SDW}.

\begin{figure}
\center
\includegraphics[width=7cm]{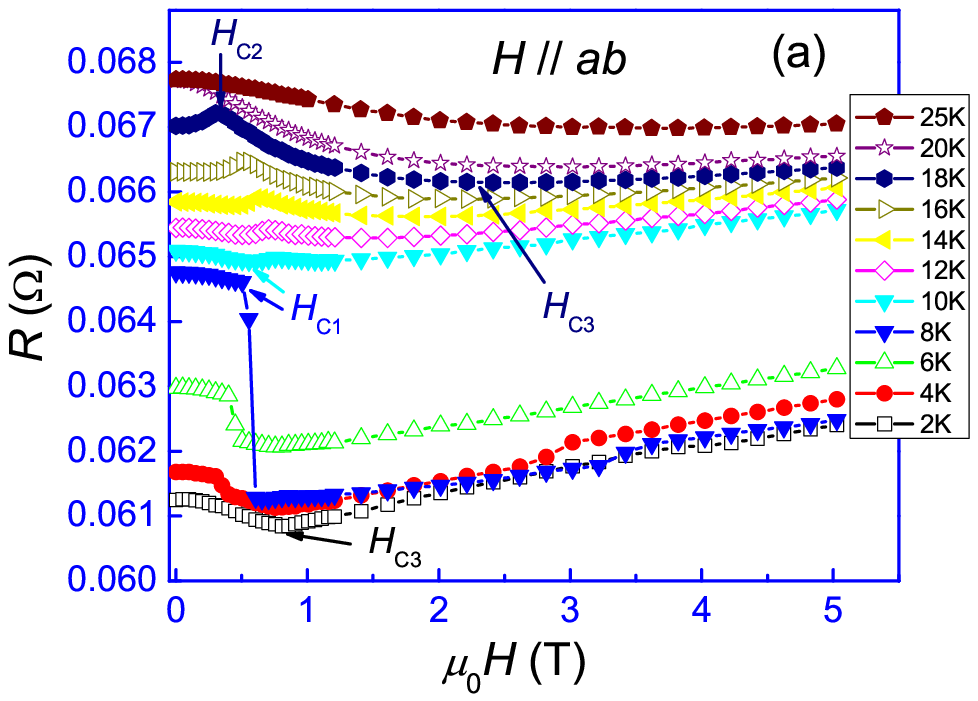}
\includegraphics[width=7cm]{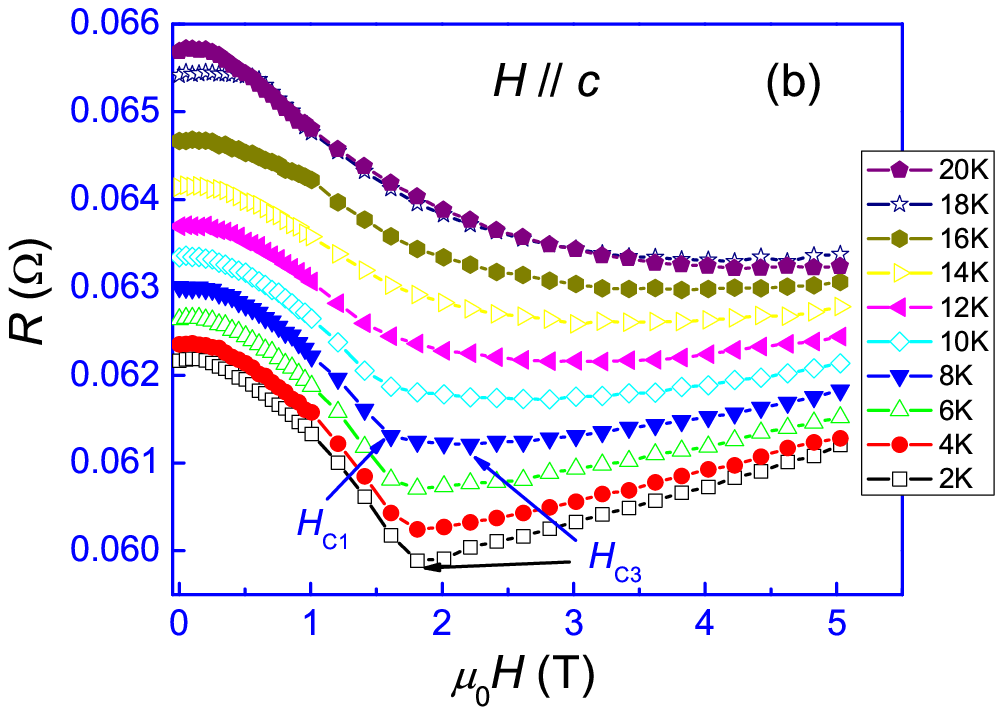}
\caption{Isothermal in-plane resistance as a function of magnetic
field for EuFe$_2$As$_2$ crystals. The applied field is
perpendicular (a) or parallel (b) to the crystallographic $c$-axis.}
\end{figure}

The $R(H)$ curves with $H\parallel ab$ are shown to be more
complicated. For clearness, the expanded $R(H)$ curves are presented
in Fig. 7. At the temperature slightly higher than $T_N$=19 K, $R$
decreases gradually with the applied field until $R$ reaches a
minimum. Since the reduction of spin-disorder scattering by external
fields leads to negative MR, on the other hand, the suppression of
Fe-SDW by the fields results in positive MR, the minimum of $R$
corresponds to the ferromagnetic alignmemt of Eu$^{2+}$ spins at
$H$=$H_{C3}$.

At the temperature range of 10 K $ \leq T \leq$ 18 K, $R(H)$ shows a
peak below $\mu_{0}H_{\parallel ab}$=1.0 T. This high MR at $H_{C2}$
suggests the spin disorder state (paramagnetic) of Eu$^{2+}$ spins.
Because the peak corresponds to the centre of magnetic hysteresis in
the $M(H)$ curves, the increase of $R$ below $H_{C2}$ is probably
due to the destruction of ferromagnetic-like order of Eu$^{2+}$
spins. On the other hand, the spin-reorientation of AFM phase causes
the decrease of $R$. Therefore, another minimum of $R$ appears at
$H_{C1}$. For 8 K $ \leq T \leq$ 4 K, a sharp drop in $R$ was
observed, which is due to the formation of a certain AFM
configuration like we proposed above.

\begin{figure}
\center
\includegraphics[width=7cm]{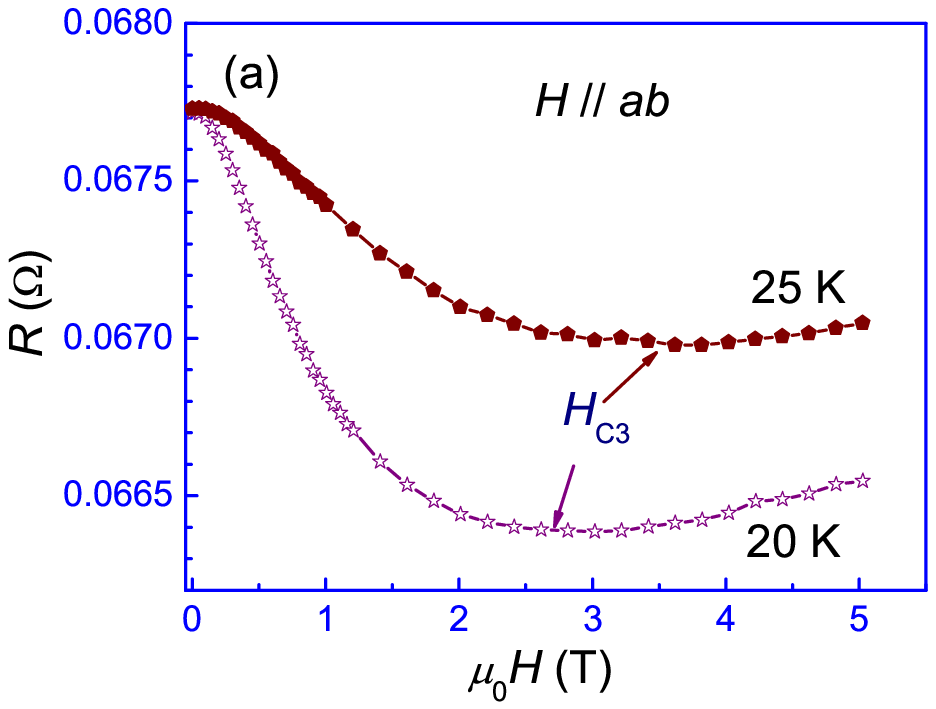}
\includegraphics[width=7cm]{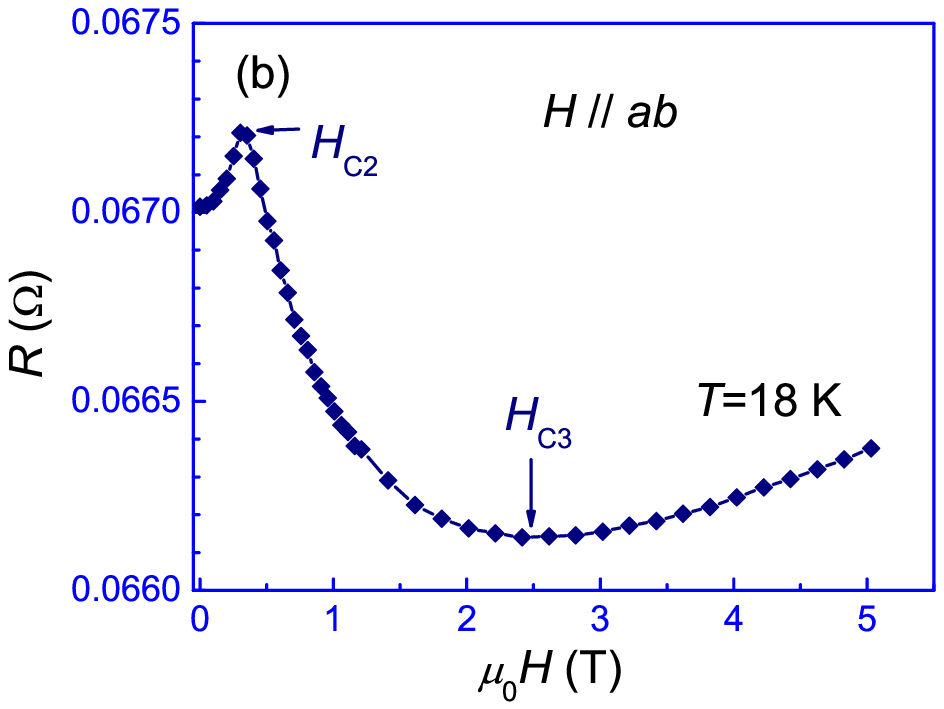}
\includegraphics[width=7cm]{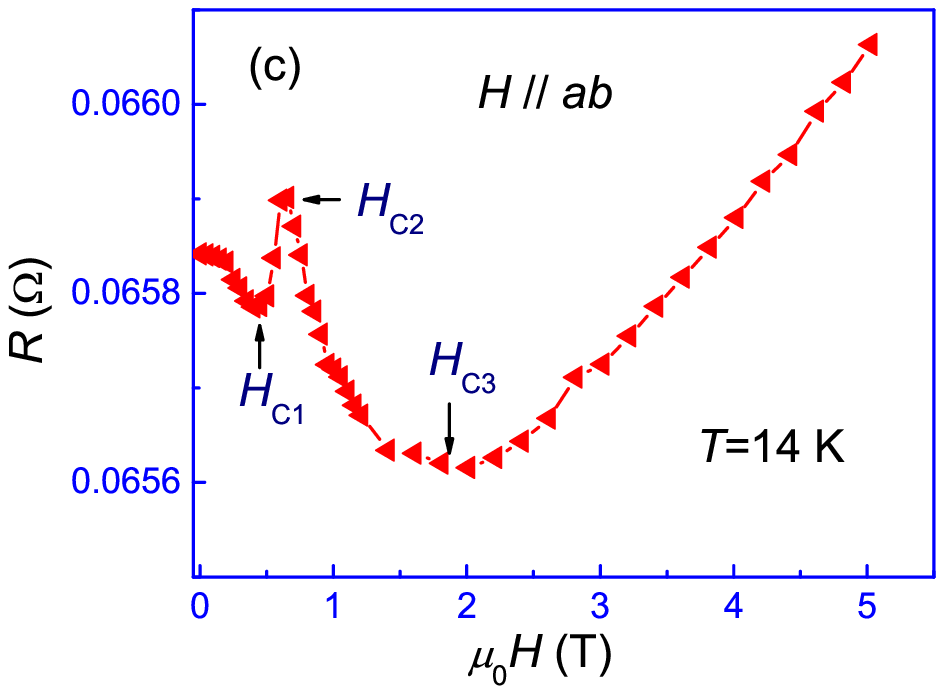}
\includegraphics[width=7cm]{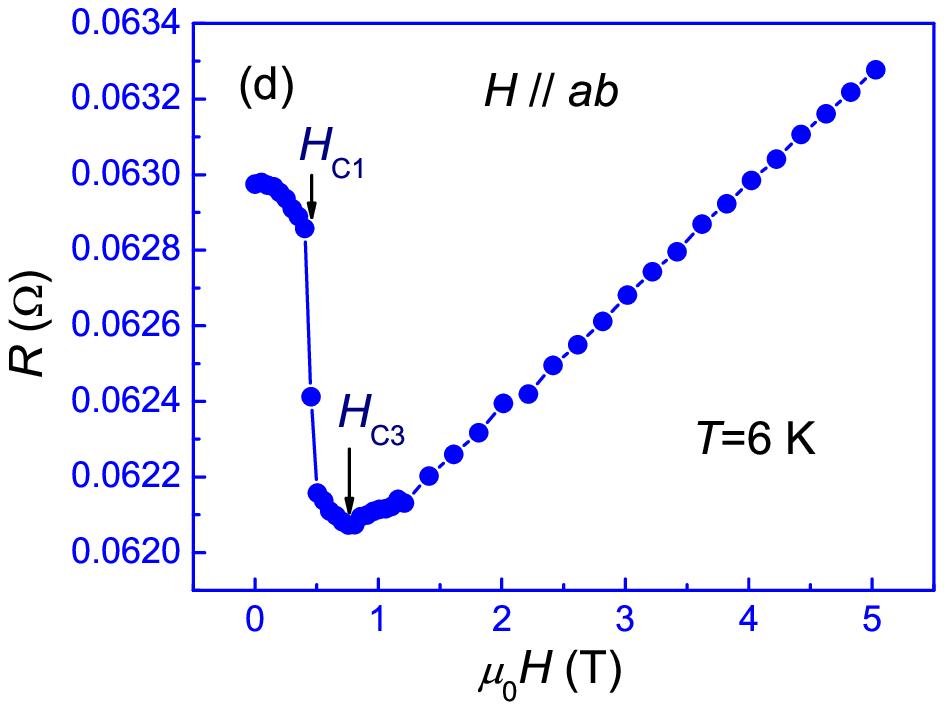}
\caption{Expanded $R_{ab}(H)$ curves with magnetic field parallel to
$ab$ planes at some representative temperatures for EuFe$_2$As$_2$
crystals. }
\end{figure}

The above data allow us to draw magnetic phase diagrams, as shown in
Fig. 8. For $H\parallel c$, one sees an AFM region at low
temperatures and low fields, in which spin re-orientation dominates.
Stronger fields lead to ferromagnetic (FM) state showing saturated
magnetic moments. The other region is paramagnetic (PM) at elevated
temperatures in which the Eu$^{2+}$ moments are aligned to some
extent by the external fields. For $H\parallel ab$, the external
fields lead to spin reversal as well as spin re-orientation. Apart
from AFM, FM and PM phases, there is an additional MM region.

\begin{figure}
\center
\includegraphics[width=7cm]{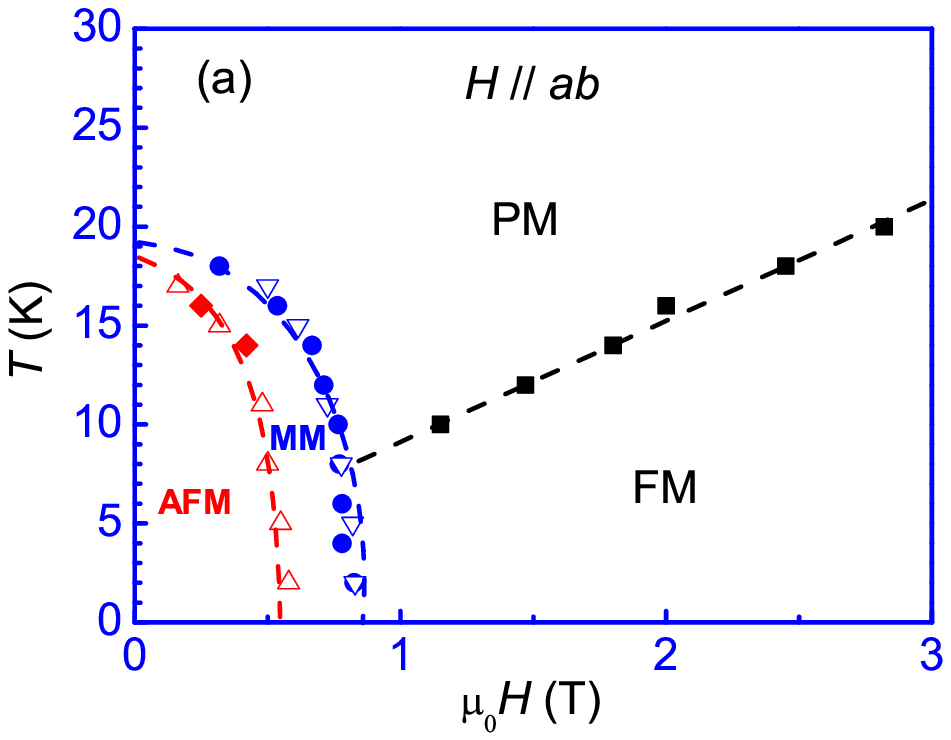}
\includegraphics[width=7cm]{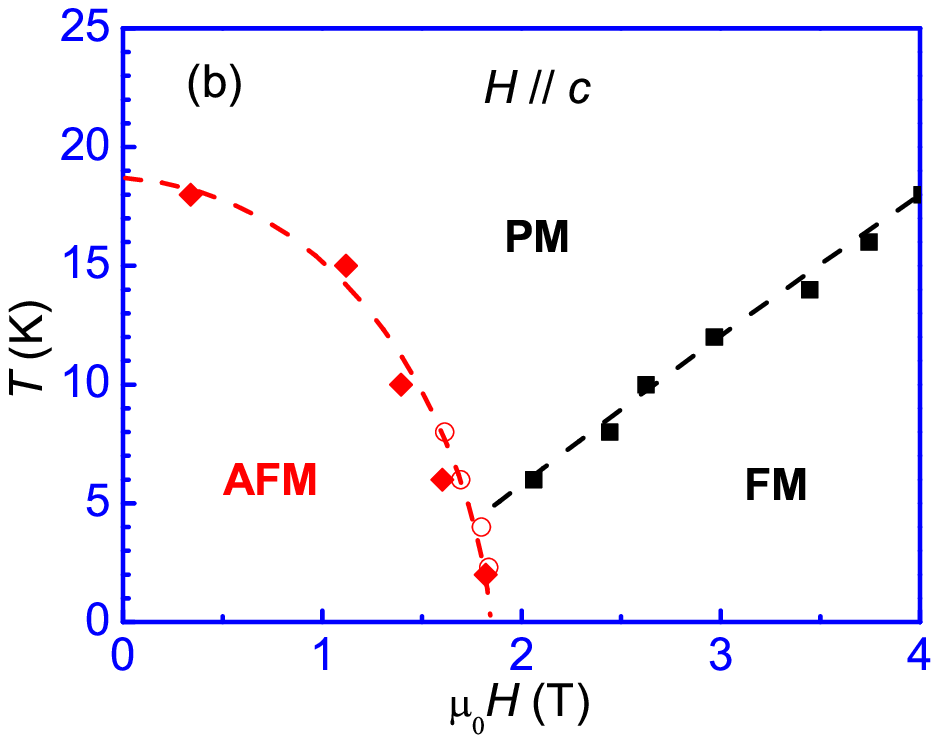}
\caption{Tentative magnetic phase diagrams for EuFe$_2$As$_2$. The
applied field is perpendicular (a) or parallel (b) to the
crystallographic $c$-axis. The open and filled symbols are from
$M(H)$ and $R(H)$ measurements, respectively. The dashed lines are
guides to the eyes.}
\end{figure}

\section{Conclusion}
To summarize, the property of AFM order of Eu$^{2+}$ spins and the
evolution of the magnetic ordering under various magnetic fields
were studied by the measurements of magnetization and
magnetoresistance using EuFe$_2$As$_2$ single crystal samples. The
result suggests that the magnetic structure for Eu$^{2+}$ spins is
of $A$-type. Under external magnetic fields with $H\parallel ab$ or
$H\parallel c$, the Eu$^{2+}$ moments undergo spin-reorientation
and/or spin-reversal transition depending on the relative
orientations between Eu$^{2+}$ moments and magnetic field. The
magnetoresistance reflects the charge carrier scattering by the
Eu$^{2+}$ moments. The electrical conduction of FeAs layers was
found to be related to the magnetic state of Eu layers. Our
preliminary result of Ni-doping\cite{Ren-Ni} in EuFe$_2$As$_2$
suggests that the magnetic state of Eu layers even influences the
appearance of superconductivity.

\section*{Acknowledgement}
This work is supported by the National Basic Research Program of
China (Contracts Nos. 2006CB601003 and 2007CB925001) and the PCSIRT
of the Ministry of Education of China (Contract No. IRT0754).

\end{document}